\documentclass[journal,10pt]{IEEEtran}
\usepackage[utf8]{inputenc}
\usepackage{amsmath,amsthm,amssymb,epsfig,mathrsfs,mathtools,cite,afterpage,color,cancel}
\usepackage[normalem]{ulem}
\usepackage{tikz,pgfplots,pgfplotstable}
\usetikzlibrary{shapes,arrows,trees}
\usetikzlibrary{decorations.pathreplacing}
\usepackage{verbatim}
\usepackage[linktoc=none]{hyperref}
\newtheorem{theorem}{Theorem}
\newtheorem{lemma}{Lemma}
\newtheorem{definition}{Definition}
\newtheorem{remarks}{Remarks}[section]
\newtheorem{corollary}{Corollary}

\def\DEFCOMPRESS{0}

    \tikzstyle{block} = [draw, fill=blue!20, rectangle, minimum height=1.5em,minimum width=2em,node distance=2.2cm]
    \tikzstyle{sum} = [draw, fill=blue!20, circle, node distance=2.2cm]
    \tikzstyle{input} = [coordinate]
    \tikzstyle{output} = [coordinate]
    \tikzstyle{pinstyle} = [pin edge={to-,thin,black}]
    \tikzstyle{connector} = [->,thick,above]
\title{
 A General Characterization of Sync Word for Asynchronous Communication
}
\author{
\IEEEauthorblockN{Sundaram R M, Devendra Jalihal, Venkatesh Ramaiyan}\\
\IEEEauthorblockA{Indian Institute of Technology Madras, Chennai 600036, India.
\\Email: sundaram.rm@gmail.com, \{dj, rvenkat\}@ee.iitm.ac.in
}
}

\begin{document}
\markboth{\tiny This work has been submitted to the IEEE for possible publication. Copyright may be
transferred without notice, after which this version may no longer be accessible.}{}
\maketitle 
\begin{abstract}
We study a problem of sequential frame synchronization
for a frame transmitted uniformly in $A$ slots. For a discrete memoryless
channel (DMC), Venkat Chandar et al showed in \cite{Chandar2008} 
that the frame length $N$ must scale with $A$ as $e^{N \alpha(Q)} > A$
for the frame synchronization error to go to zero (asymptotically with $A$). Here, $Q$ denotes the
transition probabilities of the DMC and $\alpha(Q)$, defined as the synchronization threshold,
characterizes the scaling needed of $N$ for asymptotic error free frame synchronization.
We show that the asynchronous communication framework permits a natural tradeoff between the
sync frame length $N$ and the channel (usually parameterised by the input).
For an AWGN channel, we study this tradeoff between the sync frame length $N$ and the input
symbol power $P$ and characterise the scaling needed of the 
sync frame energy $E = N P$ for optimal frame synchronisation.
\end{abstract}

\section{Introduction}
Frame synchronization generally concerns the problem of identifying the sync word 
imbedded in a continuous stream of data (see e.g., \cite{Massey1972}).
The problem of detecting and decoding frames transmitted sporadically,
possibly due to low information rate, is a subject of asynchronous communication.
The objective of an asynchronous communication
system could be, for example, to detect and decode a single frame transmitted at some random time
and there may be no transmission before or after the frame { (see e.g., \cite{Tchamkerten2009})}.

The asynchronous communication setup has been discussed in earlier works
such as \cite{Massey1972} and \cite{Mehlan1993}, but the interest
has increased in recent times with emerging applications in wireless sensor networks and the Internet of Things (IoT).
In wireless sensor and actor networks (see e.g., \cite{Akyildiz2004} and \cite{Akyildiz2002}),
the participating nodes would report a measurement or an event to the fusion centre at random epochs.
The nodes may need to transmit few bytes of data to the fusion centre over a
relatively large time frame, e.g., a single packet possibly in an hour or even in a day.
Also, in frameworks such as IoT \cite{IoT_Atzori2010}, the nodes
may report measurements sporadically leading to an asynchronous communication framework. 
However, the constraints on power may be less stringent in IoT than in wireless sensor networks.
Characterisation of the communication overheads (e.g., synchronisation overheads) needed in such set-ups is crucial
for optimal network design and operation.

\subsubsection*{Related Literature}
Earlier works on frame synchronization, such as \cite{Massey1972} and \cite{Lui1987}, used the maximum-likelihood (ML) criteria for periodically occurring  sync words. For aperiodic sync words, hypothesis testing (sequential frame sync) was preferred in works such as \cite{Corazza}, \cite{Chiani2006} and \cite{Suwan}. For the asynchronous set-up (one-shot frame sync), both ML criteria (e.g., \cite{Mehlan1993}) and hypothesis testing (e.g., \cite{Ramakrishnan} and \cite{Liang}) have been studied. These works focus only on the design and performance of receivers for a sync word designed independently.

For the asynchronous set-up, Chandar et al \cite{Chandar2008} characterized the optimum system performance considering sync word and receiver design  jointly. They study a problem of sequential frame synchronization 
for a frame transmitted randomly and uniformly in an interval of known size. For a discrete memory-less channel,
they identified a synchronisation threshold that characterises the sync frame length needed for
asymptotic error-free frame synchronisation.
In \cite{Tchamkerten2009}, following \cite{Chandar2008}, a
framework for communication in an asynchronous set up was proposed and
achievable trade-off between reliable communication and asynchronism was discussed.
In our work, we restrict to frame synchronization but generalise the framework presented in \cite{Chandar2008} 
to study a tradeoff between the sync frame length and the channel.
For the AWGN channel, this tradeoff permits us to characterise the scaling needed of the
sync frame energy (instead of the
sync frame length considered in \cite{Chandar2008} and \cite{Tchamkerten2009})
for optimal frame synchronisation.

%~~~~~~~~~~~~~~~~~~~~~~~~~~~~~~~~~~~~~~~~~~~~~~~~~~~~~~~~~~~~~~~~~~~~~~~~~~~~~~~~~~~~~~~~~~~~~~~~~~~~~~~~~~~~~~~~~~~~~~~~~~~~~~~~~~~~~~~~~~~~~~~~~~~~~
\section{System Set-up}
\label{sec:setup}
\begin{figure}
\centering
\begin{tikzpicture}[scale=4,,font=\small]
%each grid box is 0.25 width, .25 height
	\clip(-0.1,0.40) rectangle(2.1,.75);
%	\draw[step=0.25cm, gray, dotted] (-1,-1) grid (3,3);
	\foreach \i in {0,...,20} {
	  \draw[very thin] (\i/10,.5-0.02) -- (\i/10,.5+.02);
	  }
	\draw (.0,.5) -- (2,.5);
	\foreach \i in {1,...,4} {
	 \path (\i/10,0.5) node [above] {$\cdotp$};
	}
	\draw [decorate,decoration={brace}]
(1/10,0.5+0.1) -- (4/10,0.5+0.1) node [midway,above]{ $x(0)$'s};

	\path (0/10,0.5) node [below] {$0$};
	\path (20/10,0.5) node [below] {$A$};
	\path (5/10,0.5) node [above] {$s_1$};
	\path (5/10,0.5) node [above] {$s_1$};
	\path (5/10,0.5) node [below] {$v$};
	\path (6/10,0.5) node [above] {$s_2$};
	\path (7/10,0.5) node [above] {$\cdotp$};
	\path (8/10,0.5) node [above] {$\cdotp$};
	\path (9/10,0.5) node [above] {$s_N$};
%	\path (9/10,0.5) node [above] {$\tau_N$};
	\draw [thick, decorate,decoration={brace}]
(5/10,0.5+.1) -- (9/10,0.5+0.1) node [midway,above]{ $\mathbf{s}^N$};

	\foreach \i in {10,...,20} {
	 \path (\i/10,0.5) node [above] {$\cdotp$};
	}
	\draw [decorate,decoration={brace}]
(10/10,0.5+.1) -- (20/10,0.5+0.1) node [midway,above]{  $x(0)$'s};
\end{tikzpicture}
\caption{A discrete-time asynchronous communication model. A sync packet $\mathbf{s}^N = (s_1,\cdots,s_N)$
is transmitted at some random time $v \sim U\{1,A\}$.
The channel input in slots other than $\{v, \cdots, v+N-1\}$
is assumed to be $x(0)$.
}
\label{fig_async}
\end{figure}
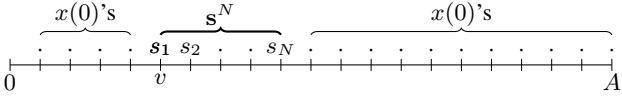

%\subsection{Asynchronous Framework}
The problem set-up is illustrated in Figure~\ref{fig_async}.
We consider discrete-time communication between a transmitter
and a receiver over a discrete memory-less channel.
The discrete memory-less channel is characterized by finite input and output alphabet sets ${\mathcal X}$ and
${\mathcal Y}$ respectively, and transition probabilities $Q(y|x)$ defined for all
$x \in {\mathcal X}$ and $y \in {\mathcal Y}$.

A sync packet $\mathbf{s}^N = (s_1, \cdots, s_N)$ of length $N$ symbols
($s_i \in {\mathcal X}$ for all $i = 1,\cdots,N$)
is transmitted at some random time, $v$, distributed uniformly in $\{1,2, \cdots, A\}$, where $A$ is assumed known.
The transmission occupies slots $\{v, v+1, \cdots, v + N -1\}$ as illustrated in Figure~\ref{fig_async},
i.e., $x_n = s_{n-v+1}$ for $n \in \{v, \cdots, v + N - 1\}$,
and, we assume that the channel input in slots other than $\{v, v+1, \cdots, v + N -1\}$ is
$x(0)$ (where $x(0) \in {\mathcal X}$ and could represent zero input).
The distribution of the channel output, $\{y_n\}$,
conditioned on the random transmission time $v$ and the
sync sequence ${\mathbf s}^N$, is
$Q(\cdot|s_{n-v+1})$ for $n \in \{v, v+1, \cdots, v+N-1\}$ and $Q(\cdot|x(0))$ otherwise.

The receiver seeks to identify the location of the sync packet $v$ from the channel output $\{y_n\}$.
Let $\hat{v}$ be
an estimate of $v$. Then, the error event
is represented as $\{\hat{v} \neq v\}$ and the associated probability of error in frame synchronization
would be ${\mathsf P}(\{\hat{v} \neq v\})$.
We are interested in characterizing the sync sequence ${\mathbf s}^N$ needed for error-free
frame synchronization as $A$ tends to infinity.
In this paper, we assume that the receiver employs a sequential decoder to detect the sync frame.
In particular, we assume that the decision $\hat{v} = t$ depends only on the output sequence up to time $t+N-1$,
i.e., $\{y_1, \cdots, y_t, \cdots , y_{t+N-1}\}$.

In \cite{Chandar2008}, Chandar et al identify a synchronization threshold that characterizes
the sync frame length needed for asynchronous optimal 
frame synchronisation.
\begin{definition}[from \cite{Chandar2008}]
\label{def:async}
Let $A = e^{N \alpha}$ denote the uncertain interval length for a given sync frame length
$N$ and a constant $\alpha$. An asynchronism exponent $\alpha$ is said to be achievable
if there exists a sequence of pairs, sync pattern and sequential decoder $( \mathbf{s}^N,\hat{v} )$, for
all $N \geq 1$, such that
\[ {\mathsf P}(\{ \hat{v} \neq v \}) \rightarrow 0 \mbox{\ \ as \ } {A \rightarrow \infty} \]
The synchronization threshold for the DMC, denoted as $\alpha(Q)$, is defined as the
supremum of the set of achievable asynchronism exponents.
\end{definition}
In \cite{Chandar2008},
the synchronization threshold for the discrete memory-less channel was shown to be
\begin{equation}
\alpha(Q) = \max_{x \in {\mathcal X}} D( Q(\cdot|x) \| Q(\cdot|x(0)) ) \label{eqn:alphadef}
\end{equation}
where $D( Q(\cdot|x) \| Q(\cdot|x(0)) )$ is the Kullback-Leibler distance between $Q(\cdot|x)$
and $Q(\cdot|x(0))$.
The authors also
provide a construction of sync sequence $\mathbf{s}^N$
entirely with two symbols, $x(0)$ and $x(1)$, where
\begin{equation}
 x(1) := \underset{x \in {\mathcal X}}{\arg\max\,}D( Q(\cdot|x) \| Q(\cdot|x(0)) ) \label{eqn:x1}
\end{equation}
and show asymptotic error-free frame synchronization with a sequential
joint typicality decoder (see section~\ref{sec:generalization} or \cite{Chandar2008} for details).
In our work, we generalise the above setup and study a tradeoff between
the sync frame length and channel parameters.

%~~~~~~~~~~~~~~~~~~~~~~~~~~~~~~~~~~~~~~~~~~~~~~~~~~~~~~~~~~~~~~~~~~~~~~~~~~~~~~~~~~~~~~~~~~~~~~~~~~~~~~~~~~~~~~~~~~~~~~~~~~~~~~~~~~~~~~~~~~~~~~~~~~~~~
\section{Motivation}
The synchronisation threshold for an AWGN channel with noise power $\sigma^2$ and input symbol power $P$
can be shown to be $\frac{P}{2 \sigma^2}$ (see \cite{Chandar2008}). 
Then, we know that the sync frame length $N$ must scale as
$e^{N \frac{P}{2 \sigma^2}} > A$ for optimal frame synchronization.
Note that this also implies a necessary scaling of the sync frame energy $E = N P$, 
i.e., $e^{N P \frac{1}{2 \sigma^2}} > A$.
This observation motivates us to study the tradeoff between the sync frame length $N$ and the
channel (and input) parameters for optimal frame synchronisation. 
In Section~\ref{sec:generalization}, for a DMC, we first present a general framework for asynchronous frame
synchronisation and then study a tradeoff between $N$ and $\alpha(Q)$.
In Section~\ref{sec:awgn}, for the AWGN channel, we discuss
the tradeoff between the sync frame length $N$ and the input symbol power $P$
and characterise the scaling needed of the sync frame energy $E = N P$ for error-free
frame synchronization.

Chandar et al \cite{Chandar2008} studied the sequential frame synchronisation problem for a
fixed $Q$ (and $\alpha(Q)$) and as a function of the sync frame length $N$ only. Also,
in \cite{Chandar2008}, the
setup and the proof based on the joint typicality of input-output sequences
requires the sync frame length $N$ to scale to infinity. In our work, we generalise the framework and study
a tradeoff between the sync frame length $N$ and channel parameters and study the case of finite sync frame
length as well.

\ifnum\DEFCOMPRESS=1
In order to conserve space, the proofs of some lemmas and corollaries are provided in \cite{Sundaram2016}.
\fi%\ifnum\DEFCOMPRESS=0
%~~~~~~~~~~~~~~~~~~~~~~~~~~~~~~~~~~~~~~~~~~~~~~~~~~~~~~~~~~~~~~~~~~~~~~~~~~~~~~~~~~~~~~~~~~~~~~~~~~~~~~~~~~~~~~~~~~~~~~~~~~~~~~~~~~~~~~~~~~~~~~~~~~~~~
\section{A General Framework for Asynchronous Frame Detection}
\label{sec:generalization}
We now present a framework that permits a tradeoff between the sync frame length $N$ and the
channel, represented by $\alpha(Q)$, for the system setup described in Section~\ref{sec:setup}.
Consider a sequence of triples, channel, sync word and sequential decoder,
$( \{ {\mathcal X}_A, {\mathcal Y}_A, Q_A \}, {\mathsf s}^{N_A}, \hat{v} )$
parameterized by the asynchronous interval length $A$.
Define $\alpha(Q_A)$ as
\begin{equation}
 \alpha(Q_A) = \underset{x \in {\mathcal X}_A}{\max} D( Q_A(\cdot|x) \| Q_A(\cdot|x_A(0)) ) \label{eqn:alpha}
\end{equation}
and let $ x_A(1) := \underset{x \in {\mathcal X}_A}{\arg\max\,}D( Q(\cdot|x) \| Q(\cdot|x_A(0)) )$.
The following theorem generalizes Theorem~1
in \cite{Chandar2008} and discusses
the necessary scaling needed of $N_A$ and $\alpha(Q_A)$ for asymptotic error-free frame synchronisation. 
\begin{theorem}
\label{thm:generalization}
Consider a sequence of triples, $( \{ {\mathcal X}_A, {\mathcal Y}_A, Q_A \}, {\mathbf s}^{N_A}, \hat{v} )$ 
parameterized
by the asynchronous interval length $A$. Let $N_A \rightarrow \infty$ as $A \rightarrow \infty$. 
Let $Q_A(\cdot | x_A(1)) \rightarrow Q_1^*(\cdot)$ and $Q_A(\cdot | x_A(0)) \rightarrow Q_0^*(\cdot)$ such that 
$\alpha(Q_A) \rightarrow \infty$ as $A \rightarrow \infty$.
Then, the probability of frame detection error ${\mathsf {P}}(\{\hat{v} \neq v\}) \rightarrow 0$ as
$A \rightarrow \infty$ if $e^{N_A \alpha(Q_A)} > A$. 
\end{theorem}

\begin{remarks}
\end{remarks}
\begin{enumerate}
\item Theorem~\ref{thm:generalization} characterizes the rate at which $N_A$ and $\alpha(Q_A)$ must scale with $A$
for the frame synchronisation error to tend to zero (asymptotically). In \cite{Chandar2008}, the channel was assumed
to be the same independent of $N$ or $A$. The generalisation proposed in Theorem~\ref{thm:generalization}
enables us to study the tradeoff between $N_A$ and $\alpha(Q_A)$ for supporting asynchronism.
\ifnum\DEFCOMPRESS=0
{\color{black!100}
\item For the AWGN channel, we know that $\alpha(Q_A) = \frac{P_A}{2 \sigma^2}$.
Hence, $N_A \times \alpha(Q_A) \propto N_A P_A$ represents the energy of the sync packet.
Thus, the above theorem also characterizes the necessary scaling needed of the energy of the sync packet
for the frame synchronisation error to tend to zero. This observation is studied in detail in Section~\ref{sec:awgn}
of this paper.
}
\fi%\ifnum\DEFCOMPRESS=0
\end{enumerate}

	Here, we have presented only the necessary outline of the proof for Theorem~\ref{thm:generalization} as the argument
	is similar to the presentation in \cite{Chandar2008}.
	\begin{IEEEproof} 		

	\textit{Setup:}
	We consider the framework presented in Section~\ref{sec:setup} for every $A$.
	A sync packet $\mathbf{s}^{N_A}$ of length $N_A$ is transmitted at some random time $v \sim U\{1,A\}$.
	The discrete memory-less channel is characterised by finite input and output alphabet sets
	${\mathcal X}_A$ and ${\mathcal Y}_A$ respectively, and transition probabilities $Q_A(\cdot|\cdot)$
	with synchronization threshold $\alpha(Q_A)$ defined as in (\ref{eqn:alpha}).
	
	\textit{Codeword:}
	Following \cite{Chandar2008}, we consider a sync sequence ${\mathbf s}^{N_A}$ of length $N_A$ with the following
	properties.
	\begin{enumerate}
	\item Fix some large $K$. Now, find a $M_A$ such that $2^{M_A - 1} - 1 < \frac{N_A}{K} \leq 2^{M_A} - 1$ for some 
	$M_A = 1,2,\cdots$.
	Let $s_n = x_A(1)$ for $2^{M_A} - 1 < n \leq N_A$. 
%	 where $K$ is any integer such that $\lfloor{\frac{N_A}{K}} \rfloor=2^m - 1$ for some $m = 1,2,\cdots$.
	Consider a maximal-length
	shift register (MLSR) sequence $\{m_n : n = 1,2,\cdots, 2^{M_A} - 1\}$
	of length $2^{M_A} - 1$ and map it to $\{ s_n : n = 1,2,\cdots,
	2^{M_A} - 1 \}$ such that $s_n = x_A(1)$ if $m_n = 0$ and $s_n = x_A(0)$ if $m_n = 1$.
	 \item The sync sequence thus obtained, ${\mathbf s}^{N_A}$, now has a Hamming distance of $\Omega\left( \frac{N_A}{2K} \right)$ with any of its shifted sequences.
	\end{enumerate}

	\textit{Decoder:}	
	We consider a simple version of the sequential joint typicality decoder for the problem setup.
	In \cite{Chandar2008}, at every time $t+N_A-1$, the decoder computes 
	the empirical joint distribution $\hat{\mathsf P}$ of the sync word (the channel input of length $N_A$)
	and the output symbols in the previous $N_A$ slots, i.e., $\{ y_{t}, \cdots, y_{t+N_A-1}\}$. 
	Whereas, we restrict our attention to those positions in the sync word 
	where we transmit symbol $x_A(1)$ and only compute
	\begin{equation*}
	\mathsf{\hat{P}}(x_A(1),y) = \frac{\mathsf{N}(x_A(1),y)}{N_A^1}, \text{ for all } y \in \mathcal{Y}
	\end{equation*}
	where, $N_A^1$ denotes the number of occurrences of $x_A(1)$ in the sync word and
	$\mathsf{N}(x_A(1),y)$ denotes the number of joint occurrences of $(x_A(1),y)$ in the sync code word and the channel output.
	We note that $N_A^1 = \Omega\left( N_A \left(1 - \frac{1}{K} \right) \right)$. 

    If the empirical distribution is close enough to the expected distribution 	$Q_A(\cdot | x_A(1))$, i.e., if $| \hat{\mathsf P}(\cdot) - 
	Q_A(\cdot|x_A(1))| < \mu$ for some fixed $\mu > 0$, then, the decoder declares $\hat{v} = t$. We have assumed that
	$Q_A(\cdot | x_A(1)) \rightarrow Q_1^* (\cdot)$ and hence, we make a simplifying assumption and declare $\hat{v} = t$ 
	only when $| \hat{\mathsf P} - Q_1^*| < \mu$.
\ifnum\DEFCOMPRESS=0
	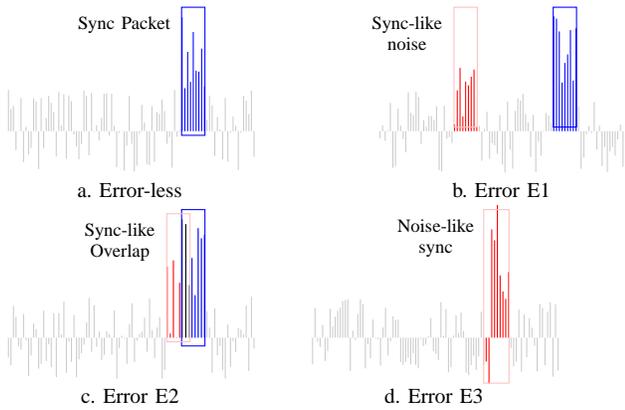
\begin{figure}
	%\includegraphics[width=.5\textwidth,clip,trim=25 25 25 25]{fading_perr_alpha_math.png}
	%shades of black for red=40,blue=70,pink=30,gray=20
	\def \cgray{black!20}
	\def \cpink{pink}
	\def \cred{red}
	\def \cblue{blue}
	\begin{tikzpicture}[scale=1,font=\scriptsize]
	\begin{axis}
	[	
	    axis y line=none,
	    axis x line=none,    
	    ymin=-1.2, ymax=3.2,
	%    group style={group size=2 by 2},
	    height = 4 cm, width = 5.5 cm,    
		title style={at={(axis cs:\xmid,-2.2)}},title=a. Error-less,font=\footnotesize,    
	]
		\def \xend{100}
		\def \xmid{50}
		\def \xbeg{1}
		\def \pbeg{71}
		\def \pend{80}
	    \addplot[ycomb,\cgray,samples=\pbeg-1-\xbeg,domain=\xbeg:\pbeg-1] {rand};
	    \addplot[ycomb,\cblue,samples=\pend-\pbeg,domain=\pbeg:\pend] {2+rand};
	    \addplot[ycomb,\cgray,samples=\xend-\pend-1,domain=\pend+1:\xend] {rand};
	   	\pgfmathsetmacro{\ya}{\pbeg-.2}
	   	\pgfmathsetmacro{\yb}{\pend+.2}
	    \draw [\cblue] (axis cs:\ya,-0.1) rectangle (axis cs:\yb,3);
	   	\pgfmathsetmacro{\yn}{\pbeg-1}
	    \node[below left , black,align=center,font=\scriptsize] at (axis cs:\yn,3) {Sync Packet};
	 \end{axis};
	 \end{tikzpicture}
	\begin{tikzpicture}[scale=1,font=\scriptsize]
	\begin{axis}
	[	
	    axis y line=none,
	    axis x line=none,    
	    ymin=-1.2, ymax=3.2,
	%    group style={group size=2 by 2},
	    height = 4 cm, width = 5.5 cm,    
		title style={at={(axis cs:\xmid,-2.2)}},title=b. Error E1,font=\footnotesize,    
	]
		\def \xend{100}
		\def \xmid{50}
		\def \xbeg{1}
		\def \pbeg{71}
		\def \pend{80}
		\def \nbeg{31}
		\def \nend{40}
	    \addplot[ycomb,\cgray,samples=\nbeg-1-\xbeg,domain=\xbeg:\nbeg-1] {rand};
	    \addplot[ycomb,\cred,samples=\nend-\nbeg,domain=\nbeg:\nend] {1+rand};
	    \addplot[ycomb,\cgray,samples=\pbeg-1-\nend-1,domain=\nend+1:\pbeg-1] {rand};
	    \addplot[ycomb,\cblue,samples=\pend-\pbeg,domain=\pbeg:\pend] {2+rand};
	    \addplot[ycomb,\cgray,samples=\xend-\pend-1,domain=\pend+1:\xend] {rand};
	   	\pgfmathsetmacro{\ya}{\pbeg-.2}
	   	\pgfmathsetmacro{\yb}{\pend+.2}
	    \draw [\cblue] (axis cs:\ya,0.1) rectangle (axis cs:\yb,3);
	   	\pgfmathsetmacro{\na}{\nbeg-.2}
	   	\pgfmathsetmacro{\nb}{\nend+.2}
	    \draw [\cpink] (axis cs:\na,0.1) rectangle (axis cs:\nb,3);
	   	\pgfmathsetmacro{\nc}{\nbeg-1}
	    \node[below left , black,align=center,font=\scriptsize]  at (axis cs:\nc,3)
		    {Sync-like\\noise};
	 \end{axis};
	 \end{tikzpicture}
	\begin{tikzpicture}[scale=1,font=\scriptsize]
	\begin{axis}
	[	
	    axis y line=none,
	    axis x line=none,    
	    ymin=-1.2, ymax=3.2,
	%    group style={group size=2 by 2},
	    height = 4 cm, width = 5.5 cm,    
	 title style={at={(axis cs:\xmid,-2.2)}},title=c. Error E2,font=\footnotesize,   
	]
		\def \xend{100}
		\def \xmid{50}
		\def \xbeg{1}
		\def \pbeg{71}
		\def \pend{80}
		\def \nbeg{65}
		\def \nend{74}
	    \addplot[ycomb,\cgray,samples=\nbeg-1-\xbeg,domain=\xbeg:\nbeg-1] {rand};
	    \addplot[ycomb,\cred,fill=\cpink,samples=\pbeg-1-\nbeg,domain=\nbeg:\pbeg-1] {1+rand};
	    \addplot[ycomb,black,fill=\cgray,samples=\nend-\pbeg,domain=\pbeg:\nend] {2+rand};
	    \addplot[ycomb,\cblue,fill=\cblue,samples=\pend-\nend-1,domain=\nend+1:\pend] {2+rand};
	    \addplot[ycomb,\cgray,samples=\xend-\pend-1,domain=\pend+1:\xend] {rand};
	   	\pgfmathsetmacro{\ya}{\pbeg-.2}
	   	\pgfmathsetmacro{\yb}{\pend+.2}
	    \draw [\cblue] (axis cs:\ya,-0.2) rectangle (axis cs:\yb,3.1);
	   	\pgfmathsetmacro{\na}{\nbeg-.2}
	   	\pgfmathsetmacro{\nb}{\nend+.2}
	    \draw [\cpink] (axis cs:\na,-0.1) rectangle (axis cs:\nb,3);
	   	\pgfmathsetmacro{\nc}{\nbeg-1}
	   	\node[below left , black,font=\scriptsize,align=center]  at (axis cs:\nc,3)
	    	{Sync-like\\Overlap};
	 \end{axis};
	 \end{tikzpicture}
	\begin{tikzpicture}[scale=1,font=\scriptsize]
	\begin{axis}
	[	
	    axis y line=none,
	    axis x line=none,    
	    ymin=-1.2, ymax=3.2,
	%    group style={group size=2 by 2},
	    height = 4 cm, width = 5.5 cm,    
	title style={at={(axis cs:\xmid,-2.2)}},title=d. Error E3,font=\footnotesize,    
	]
		\def \xend{100}
		\def \xmid{50}
		\def \xbeg{1}
		\def \pbeg{71}
		\def \pend{80}
	    \addplot[ycomb,\cgray,samples=\pbeg-1-\xbeg,domain=\xbeg:\pbeg-1] {rand};
	    \addplot[ycomb,\cred,samples=\pend-\pbeg,domain=\pbeg:\pend] {1+3*rand};
	    \addplot[ycomb,\cgray,samples=\xend-\pend-1,domain=\pend+1:\xend] {rand};
	   	\pgfmathsetmacro{\ya}{\pbeg-.2}
	   	\pgfmathsetmacro{\yb}{\pend+.2}
	    \draw [\cpink] (axis cs:\ya,-1.1) rectangle (axis cs:\yb,3.1); 
	   	\pgfmathsetmacro{\ya}{\pbeg-1}
	    \node[below left , black,align=center,font=\scriptsize] at (axis cs:\ya,3.1)
	      {Noise-like \\sync};
	 \end{axis};
	 \end{tikzpicture}
	\caption{ {\color{black!100}Error events in sequential frame synchronization problem.}	}
	\label{fig:pkt_errors}
	\end{figure}
\fi%\ifnum\DEFCOMPRESS=0

	\textit{Error event:}
The failure to detect the exact instance of sync word transmission, i.e., the error event $\{ \hat{v} \neq v \}$, can be partitioned as given below 
\ifnum\DEFCOMPRESS=0
and as shown in Figure~\ref{fig:pkt_errors}
\fi%\ifnum\DEFCOMPRESS=0
.
	\begin{itemize}
	    \item $E_1 : \hat{v} \in  \{1, \cdots, v-N_A \} \cup \{v+1, \cdots, A\}$. This corresponds to the event that the output symbols generated entirely by the zero input $x_A(0)$ is jointly typical.
	    \item $E_2 : \hat{v} \in \{v-N_A+1, \cdots, v-1\}$. This corresponds to the event that the output symbols generated partially by $x_A(0)$ and sync word is jointly typical.
		\item $E_3 : \hat{v} \notin \{ v \}$. This corresponds to the event that the output symbols generated by the sync word is not jointly typical. 
	\end{itemize}
	%We note that the event $E_1 \cup E_2$ does not contain the event $E_3$ as we consider sequential	frame detection. 
	In detection terminology, $E_1$ and $E_2$ both constitute false alarm due to noise emulation of sync word 
	and $E_3$ is missed detection.

	\textit{Performance Evaluation:}
	Using a union bound, we can upper bound the probability of error in frame synchronisation as
	\[ {\mathsf P}(\{ \hat{v} \neq v \}) \leq {\mathsf P}(E_1) + {\mathsf P}(E_2) + {\mathsf P}(E_3) \]
	Suppose that $A = e^{\epsilon_1 \cdot N_A (\alpha(Q_A) - \epsilon_2)}$ for some $0 < \epsilon_1 < 1$ and $\epsilon_2 > 0$,
	i.e., $A < e^{N_A \alpha(Q_A)}$.
	We will now show that $P(E_1), P(E_2)$ and $P(E_3)$ tend to zero as $A \rightarrow \infty$.

	The proof follows the method of types (see \cite{Csiszar2011} and \cite{Cover1991}). 
	A false alarm event of type $E_1$ occurs at a time $t$, if an input sequence composed entirely of
	$x_A(0)$ symbols generates an output type in the set 
	$\mathcal{Q}^* = \{ Q(\cdot) : |Q(y) - Q_1^*(y)| < \mu, \forall \,y \in \mathcal{Y} \}$. 
	The probability of such an event is bounded as
	\begin{eqnarray*}
	 \mathsf{P}(E_1|t) 	&\leq& \sum_{Q \in \mathcal{Q}^*} e^{-N_A^1 D(Q\|Q_A(\cdot|x_A(0)))} \\
	 &\leq& \text{poly}(N_A^1) \times e^{-N_A^1 (\alpha(Q_A) - \delta)} 
	 \end{eqnarray*}
	 where $\delta$ is a function only of $\mu$ and is independent of $A$.
	 The probability of false alarm of type $E_1$ can now be upper bounded using a union bound (over $t$) as follows.
	 \begin{eqnarray*}
	 \mathsf{P}(E_1) 	&\leq& A \times \text{poly}(N_A^1) \times e^{-N_A^1 (\alpha(Q_A) - \delta)} 
	\end{eqnarray*}
    Substituting for $A = e^{\epsilon_1 N_A (\alpha(Q_A) - \epsilon_2)}$ and bounding $N_A^1$, we have,
	 \begin{eqnarray}
	 \mathsf{P}(E_1) \leq \text{poly}(N_A) \times e^{\epsilon_1 N_A (\alpha(Q_A) - \epsilon_2)} 
	 \ifnum\DEFCOMPRESS=0
	 \nonumber\\
	 \fi%\ifnum\DEFCOMPRESS=0
	 						\times e^{- N_A \left(1 - \frac{1}{K}\right) (\alpha(Q_A) - \delta)} \label{eqn:pe1}
	\end{eqnarray}
	For large $K$ and small $\delta$ (with an appropriate choice of $\mu$), we have,
	$\mathsf{P}(E_1) \rightarrow 0$ as $A \rightarrow \infty$ (i.e., as $N_A \rightarrow \infty$ or as $\alpha(Q_A) \rightarrow \infty$).

	A false alarm event of type $E_2$ occurs if an input sequence composed partially of
	$x_A(0)$ symbols and the sync word ${\mathbf s}^{N_A}$ generates an output type in the set 
	$\mathcal{Q}^*$. 
    We note that, for every transmission instant $v$, there are $N_A - 1$ possible positions that can
    lead to the error event.
    The MLSR sequence achieves a Hamming distance of $\Omega\left( \frac{N_A}{2K} \right)$ with any of its
    shifted versions and, the Hamming distance corresponding to positions where the sync word is $x_A(1)$
    is $\Omega \left( \frac{N_A}{4K} \right)$. Using similar arguments as for $E_1$, the probability of false alarm
	of type $E_2$ can now be upper bounded as 
	\begin{eqnarray}
%	 \mathsf{P}(E_2|v) 	&\leq& \sum_{t=v-{N_A}+1}^{v-1} \mathsf{P}(E_2|v,t) \nonumber \\
	 \mathsf{P}(E_2) 	&\leq& \text{poly}(N_A) \times e^{-\Omega\left( \frac{N_A}{4K} \right) (\alpha - \delta)} \label{eqn:pe2}
	\end{eqnarray}
    Here again, ${\mathsf P}(E_2) \rightarrow 0$ as $A \rightarrow \infty$ (i.e., as $N_A \rightarrow \infty$ or as $\alpha(Q_A) 
    \rightarrow \infty$). 

    For the missed detection event $E_3$, we need to evaluate the probability that an input sequence composed entirely of
    $x_A(1)$ symbols generates an output type outside the set $\mathcal{Q}^* $. Clearly, 
	\begin{eqnarray}
	 \mathsf{P}(E_3) 	&\leq& \sum_{Q \notin \mathcal{Q}^*} e^{-N_A^1 D(Q\|Q_A(\cdot|x_A(0))} \nonumber\\
	 &\leq& \text{poly}(N_A) \times e^{-N_A \left(1 - \frac{1}{K}\right) \delta^{\prime}} 
	 \end{eqnarray}
	 where $\delta^{\prime}$ is a function only of $\mu$ and is independent of $A$.
     Now, ${\mathsf P}(E_3) \rightarrow 0$ as $A \rightarrow \infty$ (i.e., as $N_A \rightarrow \infty$).
	We note that the scaling of $\alpha(Q_A)$ does not take the error probability ${\mathsf P}(E_3) \rightarrow 0$.

	Thus, we have ${\mathsf P}(\{\hat{v} \neq v\}) \rightarrow 0$ if $N_A \rightarrow \infty$ and $\alpha(Q_A) \rightarrow \infty$ such that $e^{N_A \alpha(Q_A)} > A$.
	\end{IEEEproof}

%~~~~~~~~~~~~~~~~~~~~~~~~~~~~~~~~~~~~~~~~~~~~~~~~~~~~~~~~~~~~~~~~~~~~~~~~~~~~~~~~~~~~~~~~~~~~~~~~~~~~~~~~~~~~~~~~~~~~~~~~~~~~~~~~~~~~~~~~~~~~~~~~~~~~~
\section{Trade-off in AWGN Channel}
\label{sec:awgn}
In this section, we study the application of 
Theorem~\ref{thm:generalization} to the additive white Gaussian noise channel. 
We make the following additional assumptions to define a binary input binary output DMC model for the AWGN channel.
\begin{enumerate}
\item We consider a binary input alphabet set with ${\mathcal X_A} = \{ x_A(0) = 0, x_A(1) = \sqrt{P_A} \}$
for every $A$. $P_A$ could correspond to the 
symbol power constraint and $\frac{P_A}{\sigma^2}$ would then be the SNR.
We note that it is sufficient to consider the binary input alphabet set for the frame
synchronisation problem (see Section~\ref{sec:setup} or \cite{Chandar2008} for details).
\item The received signal at time $n$ is assumed to be $x_n + w_n$, 
where $w_n$ is WGN with variance $\sigma^2$.
\item We consider a binary alphabet set ${\mathcal Y}_A$ for the output channel, i.e., 
$\mathcal{Y}_A=\{y_A(0),y_A(1)\}$ for every $A$.
In particular, we consider the following map for the AWGN channel: the output is $y_A(1)$ if
$x_n + w_n > \tau_A = a \sqrt{P_A}$ for some $0 < a < 1$ and the output is $y_A(0)$ if
$x_n + w_n \leq \tau_A$. 
\ifnum\DEFCOMPRESS=0
The binary input and binary output DMC model for the AWGN channel is illustrated
in Figure~\ref{fig_bdmc} where $\epsilon_f$ and $\epsilon_m$ denote the transition probabilities.
\fi%\ifnum\DEFCOMPRESS=0
We show in Section~\ref{sec:binary_awgn} that the two alphabet approximation for the output channel
is appropriate in the context of asynchronous frame synchronisation.
\end{enumerate}
\subsection{Binary Output DMC Model for AWGN Channel}
\label{sec:binary_awgn}
The synchronisation threshold for the AWGN channel with noise power $\sigma^2$ and input symbol power $P$ was
shown to be $\frac{P}{2 \sigma^2}$ (see \cite{Chandar2008}). The following lemma shows that the
binary input binary output model for the AWGN channel can achieve a synchronisation threshold 
arbitrarily close to $\frac{P}{2 \sigma^2}$.
\begin{lemma}
Consider the binary input binary output model for the AWGN channel
\ifnum\DEFCOMPRESS=0
 shown in Figure~\ref{fig_bdmc}
\fi%\ifnum\DEFCOMPRESS=0
.
The synchronization threshold of the DMC tends to $\frac{P}{2 \sigma^2}$ for $a \approx 1$ and as $P \rightarrow \infty$.
\ifnum\DEFCOMPRESS=1
   \qed
\fi%\ifnum\DEFCOMPRESS=0
\end{lemma}
\ifnum\DEFCOMPRESS=0
{\color{black!100}
\begin{IEEEproof}
The channel transition probabilities for the DMC are 
\[
\epsilon_f=\mathsf{P}(y_A(1)|x_A(0))=\mathsf{P}(n>a\sqrt{P}) \simeq e^{-\frac{a^2 P}{2\sigma^2}}
\]
\[
\epsilon_m=\mathsf{P}(y_A(0)|x_A(1))=\mathsf{P}(n>(1-a)\sqrt{P}) \simeq e^{-\frac{(1-a)^2 P}{2\sigma^2}}
\]
The synchronization threshold for the binary DMC is given by
\[
\alpha = (1-\epsilon_m) \log\frac{1-\epsilon_m}{\epsilon_f}+\epsilon_m \log \frac{\epsilon_m}{1-\epsilon_f}
\]
Clearly, $\epsilon_f, \epsilon_m \rightarrow 0$ as $P \rightarrow \infty$. Hence,
\begin{eqnarray*}
\alpha &\underset{P \rightarrow \infty}{\rightarrow}& -\log \epsilon_f + \epsilon_m \log \epsilon_m\\
&\simeq& \frac{a^2 P}{2\sigma^2} +  \frac{(1-a)^2 P}{2\sigma^2} e^{-\frac{(1-a)^2 P}{2\sigma^2}}\\
&\simeq& \frac{a^2 P}{2\sigma^2}
\end{eqnarray*}
Thus, for large $P$ and $a$ close to $1$, the synchronization threshold of the binary input binary output 
tends to the synchronisation threshold of the AWGN channel.
\end{IEEEproof}
}
\fi%\ifnum\DEFCOMPRESS=0
The above lemma permits us to apply the results of the Section~\ref{sec:generalization} for the AWGN channel.
\ifnum\DEFCOMPRESS=0
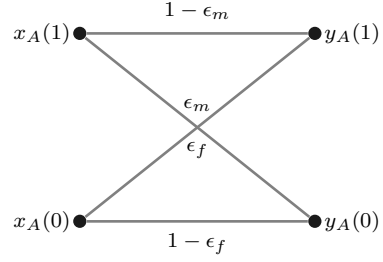
\begin{figure}
\centering
\def\layersep{2.5cm}
\begin{tikzpicture}[draw=black!50, node distance=\layersep,font=\footnotesize,scale=1.25]
    \tikzstyle{sym}=[circle,fill=black!20,minimum size=5pt,inner sep=0pt]
    \tikzstyle{xsym}=[sym, color=black!90,text=black];
    \tikzstyle{ysym}=[sym, color=black!90,text=black];
    \tikzstyle{hsym}=[sym, color=black!90,text=black];

    % Draw the input layer nodes
    \foreach \name / \y in {1,2}
    {
    % This is the same as writing \foreach \name / \y in {1/1,2/2,3/3,4/4}
        \path[yshift=.5cm] node[xsym] (I-\name) at (0,-2*\y) {};
        \node[left of=I-\name,node distance=10] {};
    }
          
    % Draw the output layer node
    \foreach \name / \y in {1,2}
    {
	 \path[yshift=0.5cm] node[ysym] (O-\name) at (1*\layersep,-2*\y) {};
	 \node[right of=O-\name,node distance=10] {};
	}

    % Connect every node in the fading  layer with the output layer
    \foreach \source in {1,2}
        \foreach \dest in {1,2}
            \path [line width=1pt] (I-\source) edge node (IO-\source\dest) {} (O-\dest);

    \node[left of=I-1,node distance=.5cm] {$x_A(1)$};
    \node[left of=I-2,node distance=.5cm] {$x_A(0)$};
    \node[right of=O-1,node distance=.5cm] {$y_A(1)$};
    \node[right of=O-2,node distance=.5cm] {$y_A(0)$};
    \node[above of=IO-11,node distance=0.3 cm] {$1-\epsilon_m$};
    \node[above of=IO-12,node distance=0.3 cm] {$\epsilon_m$};
    \node[below of=IO-21,node distance=0.3 cm] {$\epsilon_f$};
    \node[below of=IO-22,node distance=0.3 cm] {$1-\epsilon_f$};
\end{tikzpicture}
\caption{\color{black!100}A binary input binary output model for AWGN channel with transition probabilities $\epsilon_f = \mathsf{P}(y_A(1)|x_A(0)) =\mathsf{P}(n>a\sqrt{P}) $ and $\epsilon_m = \mathsf{P}(y_A(0)|x_A(1))=\mathsf{P}(n>(1-a)\sqrt{P})$.}
\label{fig_bdmc}
\end{figure}
\fi%\ifnum\DEFCOMPRESS=0
%~~~~~~~~~~~~~~~~~~~~~~~~~~~~~~~~~~~~~~~~~~~~~~~~~~~~~~~~~~~~~~~~~~~~~~~~~~~~~~~~~~~~~~~~~~~~~~~~~~~~~~~~~~~~~~~~~~~~~~~~~~~~~~~~~~~~~~~~~~~~~~~~~~~~~
\subsection{Tradeoff for the AWGN Channel}
The following corollary discusses an application of Theorem~\ref{thm:generalization} 
for the AWGN channel.
\begin{corollary}
\label{cor:energy}
Consider an AWGN channel with noise variance $\sigma^2$. 
Let $N_A$ and $P_A$ denote the sync word length and the input symbol power parameterized by
the asynchronous interval length $A$.
Let $N_A, P_A \rightarrow \infty$ as $A \rightarrow \infty$.
Then, the probability of frame detection error ${\mathsf {P}}(\{\hat{v} \neq v\}) \rightarrow 0$
if $e^{N_A P_A \frac{1}{2 \sigma^2}} > A$. 
\ifnum\DEFCOMPRESS=1
   \qed
\fi%\ifnum\DEFCOMPRESS=0
\end{corollary}
\ifnum\DEFCOMPRESS=0
{\color{black!100}
	\begin{IEEEproof}
%	From Equation~\ref{eqn:alpha_w}, 
	We know that $\alpha(Q_A) \rightarrow \frac{a^2 P_A}{2 \sigma^2} \simeq \frac{P_A}{2 \sigma^2}$ for the binary input binary output model for the AWGN channel as $P_A \rightarrow \infty$. Also, as $P_A \rightarrow \infty$, we see that $Q_A(\cdot|x_A(1)) \rightarrow (0,1)$ and $Q_A(\cdot|x_A(0)) \rightarrow (1,0)$ satisfying the assumptions. Hence, $e^{N_A \alpha(Q_A)} \rightarrow e^{N_A \frac{P_A}{2 \sigma^2}}$ as $A \rightarrow \infty$.
	From Theorem~\ref{thm:generalization}, we then have ${\mathsf {P}}(\{\hat{v} \neq v\}) \rightarrow 0$ as $A \rightarrow \infty$
	if $e^{N_A \frac{P_A}{2\sigma^2}}> A$.
\end{IEEEproof}
}
\fi%\ifnum\DEFCOMPRESS=0
%The following corollary discusses the result for the special case $N_A = 1$ (and any bounded $N_A$ as well).
%\begin{corollary}
%Consider an AWGN channel with $\alpha(Q_A) = \frac{P_A}{8 \sigma^2}$, where $\frac{P_A}{\sigma^2}$ is the
%SNR. Let $N_A = 1$ for all $A$. Then, ${\mathsf {P_A}}(\hat{v} \neq v) \rightarrow 0$ if $P_A \rightarrow \infty$ such that
%$e^{\frac{1}{8 \sigma^2} P_A} > A$.
%\end{corollary}
\begin{remarks}
\end{remarks}
\begin{enumerate}
 \item Define $E_A = N_A P_A$ as the energy of the sync packet. Then, the above corollary characterises
 the scaling necessary of the energy of the sync packet (when both $N_A$ and $P_A$ are adapted)
 for asymptotic error-free frame synchronisation.
\end{enumerate}

The following lemma extends the results of Corollary~\ref{cor:energy} for a sync word of finite length.
\begin{lemma}
 \label{lem:finiteN}
  Consider an AWGN channel with noise variance $\sigma^2$. Let $N_A$ and $P_A$ denote the sync
  word length and the input symbol power parameterised by the asynchronous interval length $A$. 
  Let $N_A = N$ for all $A$ and let $P_A \rightarrow \infty$ as $A \rightarrow \infty$.
  Then, the probability of frame detection error ${\mathsf {P}}(\left\{ \hat{v} \neq v \right\}) \rightarrow 0$ if
  $e^{N P_A \frac{1}{2 \sigma^2}} > A$. 
\ifnum\DEFCOMPRESS=1
   \qed
\fi%\ifnum\DEFCOMPRESS=0
\end{lemma}
\ifnum\DEFCOMPRESS=0
{\color{black!100}
\begin{IEEEproof}
The proof follows similar arguments as in Theorem~\ref{thm:generalization}. Here again, we seek to show that
  ${\mathsf P}(\{ \hat{v} \neq v \}) \leq \mathsf{P}(E_1)+ \mathsf{P}(E_2)+\mathsf{P}(E_3) \rightarrow 0$ under the suggested
  conditions.
  
  \textit{Codeword}:
  Since $N$ is finite, we will simplify the sync word and let it consist only of $x_A(1) = \sqrt{P_A}$ in all the positions.
  
  \textit{Decoder}:
   As the sync word comprises only of $x_A(1)$, the entire length of the sync word is used for decoding. As $P_A \rightarrow \infty$, we see that $Q_A(\cdot|x_A(1)) \rightarrow (0,1) = Q_1^*$. The decoder will declare $\hat{v} = t$ if 
   $| \hat{{\mathsf P}} - Q_1^*| < \mu$. For the finite $N$ case, we will set $\mu = \frac{1}{N}$. Then, for the choice of $\mu$,
   we have ${\mathcal Q}^* = \{ Q(\cdot) : | Q(y) - Q_1^*(y)| < \frac{1}{N}, \forall y\} = \{ (0,1) \}$.
  This implies that the decoder will declare the sync packet as received only when all the previous $N$ output symbols are decoded as $y_A(1)$.
   
  \textit{Performance Evaluation}:
The probability of false alarm of type $E_1$ for the decoder can now be upper bounded as
  \begin{eqnarray*}
  \mathsf{P}(E_1) { \leq A \times \epsilon_f^N \leq A \times e^{-N \frac{a^2 P_A}{2 \sigma^2}} }
  \end{eqnarray*}
If $A = e^{\epsilon_1 N \frac{P_A}{2 \sigma^2}} < e^{N \frac{P_A}{2 \sigma^2}}$ for some $0 < \epsilon_1 < 1$, then we have $\mathsf{P}(E_1) \rightarrow 0$ as $P_A \rightarrow \infty$ for a suitable choice of $a$.
  
The probability of false alarm of type $E_2$ can be upper bounded by considering the worst case overlap with the sync word 
and using a union bound as given below.
  \begin{eqnarray*}
  \mathsf{P}(E_2) { \leq (N-1) \epsilon_f \leq (N-1) e^{-\frac{a^2 P_A}{2 \sigma^2}}}
  \end{eqnarray*}
  Clearly, $\mathsf{P}(E_2) \rightarrow 0$ as $P_A \rightarrow \infty$.

The missed detection occurs even if one of the symbols is in error, since $\mu = \frac{1}{N}$. Thus, using a union bound, the probability of 
missed detection is upper bounded as 
  \begin{eqnarray*}
	 \mathsf{P}(E_3) { \leq N \epsilon_m \leq N e^{-\frac{(1-a)^2 P_A}{2 \sigma^2}}}
  \end{eqnarray*}
${\mathsf P}(E_3) \rightarrow 0$ as $P_A \rightarrow \infty$. Hence, ${\mathsf P}(\{ \hat{v} \neq v \}) \rightarrow 0$
as $P_A \rightarrow \infty$.
\end{IEEEproof}
}
\fi%\ifnum\DEFCOMPRESS=0

\begin{remarks}
\end{remarks}
\begin{enumerate}
  \item Let $N = 1$. The above lemma suggests that we can achieve arbitrarily low packet 
   detection error if $e^{\frac{1}{2 \sigma^2} P_A} > A$, even with a single length sync word.
   \item We note again that the proofs (in Section~\ref{sec:generalization} and in earlier references \cite{Chandar2008} and \cite{Tchamkerten2009}) 
   based on joint typicality of input-output sequences require the sync frame length $N$ to scale to infinity. In Lemma~\ref{lem:finiteN}, we 
   illustrate that asynchronous frame synchronization over an AWGN channel can be achieved with finite sync frame length as well.
\ifnum\DEFCOMPRESS=0
   \item In the proof of Theorem~\ref{thm:generalization}, for a general DMC,
   we noted that ${\mathsf P}(E_3)$ need not scale to zero
   as $\alpha(Q_A) \rightarrow \infty$. However, in the binary input binary output model for the AWGN channel 
    shown in Figure~\ref{fig_bdmc}, 
   ${\mathsf P}(E_3) \rightarrow 0$ as $\alpha(Q_A) \rightarrow \infty$. This permits us to describe an asynchronous frame
   synchronisation framework for a finite length sync word.
\fi%\ifnum\DEFCOMPRESS=0
\end{enumerate}
Motivated by the results obtained so far for AWGN channel,
\begin{enumerate}
 \item $P_A=P$, $N_A \rightarrow \infty$ as $A \rightarrow \infty$ by Chandar \cite{Chandar2008}
 \item $P_A \rightarrow \infty$, $N_A \rightarrow \infty$ as $A \rightarrow \infty$ in Corollary~\ref{cor:energy}
 \item $P_A \rightarrow \infty$, $N_A=N$ as $A \rightarrow \infty$ in Lemma~\ref{lem:finiteN}
\end{enumerate}
we can now define the synchronization threshold for the AWGN channel in terms of the sync packet energy.

\begin{lemma}
The synchronisation threshold for the AWGN channel with respect to the sync packet energy is $\frac{1}{2 \sigma^2}$.
 \qed
\end{lemma}

%~~~~~~~~~~~~~~~~~~~~~~~~~~~~~~~~~~~~~~~~~~~~~~~~~~~~~~~~~~~~~~~~~~~~~~~~~~~~~~~~~~~~~~~~~~~~~~~~~~~~~~~~~~~~~~~~~~~~~~~~~~~~~~~~~~~~~~~~~~~~~~~~~~~~~
\section{Conclusion}
In this paper, we present a general framework for asynchronous frame synchronisation
that permits a trade-off between sync word length $N$ and
channel.
The framework allowed us to characterise the synchronisation threshold for the AWGN channel
in terms of the sync frame energy (i.e., $e^{E \frac{1}{2 \sigma^2}} > A$) instead of the sync frame length.
We also observe that a finite sync word can achieve optimal frame synchronization for an AWGN channel. 
As future work, we seek to study this trade-off for wireless channel models. 

\bibliographystyle{IEEEtran}
\bibliography{bibfile_short}
\end{document}